\documentstyle[twocolumn,prl,aps]{revtex}
\begin{document}
\draft

\title{Orientational Order and Depinning of the Disordered Electron Solid}

\author{Min-Chul~Cha and H.~A.~Fertig}
\address{Department of Physics and Astronomy and Center for Computational
Sciences, University of Kentucky, Lexington, KY~~40506}

\date{\today}

\maketitle
\begin{abstract}
We study the ground state of two-dimensional classical electron
solids under the influence of modulation-doped impurities
by using a simulated annealing molecular dynamics method.
By changing the setback distance as a parameter,
we find that in the strong disorder limit the ground state configuration
contains both isolated dislocations and disclinations,
whereas in the weak disorder regime only dislocations are present.
We show, via continuum elasticity theory,
that the ground state of the lattice should be unstable against
a proliferation of free disclinations above a critical dislocation density.
Associated with this, the behavior of the threshold
electric field as a function of the setback distance changes.
\end{abstract}


\narrowtext
Since Wigner\cite{Wigner1934} proposed a solid phase of electrons,
searches for a Wigner crystal (WC) have been pursued in various systems.
In a low density region, a realization of the WC on a helium film
has been established\cite{Grimes1979}.
The WC is also expected to be formed in two-dimensional systems in 
very strong magnetic fields\cite{MIWC}.
Recently a considerable amount of experimental study has focused on the
search for this magnetically induced WC\cite{exp1,exp2}.
All of these observations have been interpreted as indirect indications
of a pinned WC.
There is also some evidence of experimental observation of the WC
at zero magnetic field in Si MOSFET\cite{Pudalov1993}.
However, disorder, which is ubiquitous, makes it difficult to interpret
the experimental observations.
For example, the value of the depinning threshold electric field
of the presumed magnetically induced WC for slightly different sample
geometries differs up two orders of magnitude\cite{exp1}.
To interpret experimental observations, therefore, it is necessary to
understand the pinning sources and estimate the pinning forces.
One aspect of this problem is that random impurities pin a WC and at
the same time deform it, introducing defects in the lattice.
Only a few studies have been devoted to clarify the
possible sources of pinning\cite{Ruzin1992} in the two dimensional electron
system confined in a heterojunction.
Furthermore, the details of the relation
between the defects generated by impurities and the pinning\cite{Shi1991}
are not well understood.

In this paper, we study pinning and orientational order in
a model of the classical two-dimensional electron system.
Our calculations are most directly applicable to electrons
on a helium film\cite{Jiang1989},
although some of our basic conclusions should apply to heterojunction
systems in strong magnetic fields,
where the filling factor is quite small, so that
quantum exchange effects are unimportant\cite{Maki1983}.
Disorder is introduced by modulation-doped donors randomly located on a
plane separated by a setback distance $d$
from the two-dimensional electron plane.
Since the charged electrons and ions are interacting by a $1/r$ potential,
the effect of impurities is characterized by a dimensionless constant
$d/a_0$, where $a_0$ is the lattice constant of a perfect lattice.
It is known that arbitrarily weak disorder destroys
long-range translational order in two dimensions associated with
a crystal lattice\cite{Imry1975}.
However, we can study the {\it orientational\/}\ order of this system,
and the associated change in behavior of the depinning threshold electric field.
From numerical simulations with up to 3200 particles,
we have observed defects ---predominantly dislocations and disclinations---
generated by the impurity potential.
In the weak disorder limit (large $d/a_0$),
the ground state configuration contains a quasi-long range orientational order
\cite{Nelson1982b}({\it i.e.\/}\ hexatic phase),
and we observed no free disclinations in the system.
However, as disorder increases, isolated disclinations appear,
destroying the quasi-long range orientational order\cite{Nelson}.
We will argue below, based on continuum elasticity theory, 
that above a threshold density of dislocations, it is always energetically
favorable to create isolated disclinations.

Our principal results are summarized in Figs.~\ref{correlation}
and \ref{efield}.
Fig.~\ref{correlation} shows the orientational correlation functions
for different values of $d/a_0$.
One sees that the quasi-long range orientational order is destroyed
in the strong disorder limit.
Associated with this crossover, the behavior of the threshold electric
field is changed as shown in Fig.~\ref{efield}.
The crossover takes place approximately at $d/a_0 \approx 1.15 \pm 0.1$,
which is consistent with the vanishing of orientational order.
To qualitatively understand this behavior, it is necessary to observe
the motion of the electrons as they depin.(Details will be published
elsewhere\cite{Cha1994c}.)
We find that the electrons tend to flow along directions of the
local bond orientation;
{\it i.e.\/}, to flow along local symmetry directions of the crystal.
Since the system does not have long range orientational order,
it is necessary for electron to pass through regions of great
strain in the lattice, where the orientation changes.
These regions of strain represent bottlenecks in the electron flow.
As the disorder strength is tuned and orientational order changes
from quasi-long range to short range,
the number of bottlenecks proliferates
and there is a sharp increase in the threshold field.
We also note that the threshold electric field is very sensitive
to the setback distance, which might explain the very disparate
values of this quantity in experiments\cite{exp1}.

More explicitly, we study a system whose energy is given by
\begin{equation}
E
  =\sum_{i \ne j}{e^2 \over \epsilon |\vec r_i - \vec r_j|}
  -\sum_{ij}{e^2 \over \epsilon (|\vec r_i - \vec R_j|^2 + d^2)^{1/2}}
\end{equation}
where $\{\vec r_i\}$ are the electron configurations,
$\{\vec R_i\}$ are quenched donor configurations on a modulation-doped plane,
$d$ is the setback distance, and $\epsilon$ is a dielectric constant.
Numerically, we use a simulated annealing molecular dynamics
method\cite{Brass1989} to find
an electron configuration, $\{\vec r_i\}$, that corresponds to
the ground state or, at least, a typical low energy metastable state.
We impose periodic boundary conditions,
and use the Ewald sum technique\cite{Bonsall1977} to
handle the long range interaction in computing energies and forces.
Because our study focuses in part on the depinning properties of this system,
and it is possible that the depinning will be inhomogeneous
({\it i.e.\/}, that current paths form inside the crystal),
we must account for the fact that electrons exiting the system will need
to reenter it on the opposite side of the unit cell.
To minimize any mismatch of current patterns at the boundaries,
along the direction of the depinning field we juxtapose two square boxes whose
impurity configurations are mirror-images of each other.

We take $\epsilon=13$ in the numerical simulations,
and slowly lower the temperature from above the melting
transition down to $20{\rm mK}$
for a typical electron density $n=5.7 \times 10^{10} {\rm cm}^{-2}$.
At this temperature we measure the orientational correlation functions,
which are summarized in Fig.~\ref{correlation}.
The correlation functions can be well fitted by exponential forms in
the strong disorder limit (small $d$).
We find that the correlation lenght $\xi$ is a slowly varying function
for $d/a_0 <1$;
in the interval $1.1 < d/a_0 <1.2$, the correlation length
rises rapidly,
suggesting a possible divergence (and an associated phase transition).
However, once the correlation length exceeds our system size,
fits to either an exponential form or a power law become possible,
and it becomes difficult to precisely identify what value of $d/a_0$
would be the critical one in an infinite system.
However, based on the sharp increase in $\xi(d)$,
we suspect that it is in the vicinity of 1.2.
This estimate is also consistent with the observation of the
structure factor in Fig.~\ref{sfactor}, where the six-fold symmetry of
the orientational order\cite{Nelson1982b} appears only for the samples
with setback distance bigger than 1.1.

We also measure the threshold depinning field.
To do this, we shift the positions of the particles along a chosen direction
by steps of $0.01 a_0$\cite{Brass1989} up to 1--3 lattice constants.
After each shift 200 MD steps are taken to equilibrate the system,
and the pinning force is obtained in the next 100 MD steps
from the averaging the force on particles due to impurities.
The threshold field is determined by the maximum value of the pinning force
multiplied by the dielectric constant during the shifting process;
error bounds may be obtained by comparing the heights of the several peaks
that one sees.
The results are shown in Fig.~\ref{efield}.
We take two samples for each setback distance.
We also determine the threshold field for several samples
by gradually increasing an external field and by identifying the point
at which the currents start to flow.
The results obtained through the two methods are in good agreement.
The behavior of the threshold depinning force changes around
$d/a_0=1.15 \pm 0.1$,
which suggest that disclinations produce extra pinning of the lattice.

We now describe how free disclinations might be favored in strongly
disordered samples using standard elasticity theory\cite{Landau}.
Since the charged system requires charge neutrality at long wavelengths,
we consider a model whose elastic energy is given by
\begin{equation}
\tilde E
  ={1 \over 2}\int d^2\vec r\ [2\mu u_{ij}^2
  +\lambda (\nabla \cdot \vec u - \delta \rho(\vec r))^2]\;,
\label{elastic_energy}
\end{equation}
where $\mu$ and $\lambda$ are Lam\'e coefficients,
$\vec u(\vec r)$ are displacement vectors,
$u_{ij}=(1/2)(\partial u_i/\partial x_j + \partial u_j /x_i)$, and
$\rho (\vec r) = \rho_0 +a_0^{-2} \delta \rho(\vec r)$ is an effective in-plane
impurity density.
(Summation convention is assumed for repeated indices.)
Since the longitudinal sound velocity goes to infinity for $1/r$ interaction,
we take the limit $\lambda \to \infty$ at the end of our calculation.
This will guarantee that the electron density tracks the neutralizing
background, which is the correct physics at long wavelengths.
With defining a stress tensor
\begin{equation}
\Pi_{ij}=2 \mu u_{ij}+ \lambda \delta_{ij} (\nabla \cdot \vec u)\;,
\end{equation}
we can divide Eq.\ (\ref{elastic_energy}) into the contributions
from smooth elastic displacements and displacements related to defects
which involve singularities.
\begin{mathletters}
\begin{eqnarray}
\tilde E&=&E_0 + E^{\prime}\;, \\
E^0&=&{\lambda \over 2} \int d^2\vec r \ \delta\rho
    [\delta\rho - \nabla \cdot \vec u^0]\nonumber \\
   &=&{\lambda \over 2} \int d^2\vec r \ (\delta\rho)^2
    [1-{\lambda \over 2\mu+\lambda}]
    \to \mu \int d^2\vec r \ (\delta \rho)^2\;, \\
E^{\prime}&=&{1 \over 2} \int d^2\vec r \ [\Pi^{\prime}_{ij} u^{\prime}_{ij}
    -\lambda (\nabla \cdot \vec u^{\prime}) \delta \rho ]\;,
\end{eqnarray}
\end{mathletters}
where $\vec u^0$ and $\vec u^{\prime}$ represent the regular and the singular
parts, respectively, etc.
We note that all these quantities are well-behaved as $\lambda \to \infty$.
As with many crystal systems in the presence of disorder
\cite{Shi1991,Coppersmith1990},
we find that our electron crystal is unstable against the formation of
highly separated dislocations even in the weak disorder limits\cite{Cha1994c},
and this is certainly the case in our simulations.
As the disorder strength is increased, the density of dislocations increases.
It is instructive to consider the energy of a single disclination in
a complexion of dislocations.
We write the defect energy as
\begin{equation}
E^{\prime}=E^{\prime}_{D}+E^{\prime}_{BD}+E^{\prime}_{D\rho}\;,
\end{equation}
where $E^{\prime}_{D}, E^{\prime}_{BD}$, and $E^{\prime}_{D\rho}$ are
energy of the disclination, the coupling energy between disclination and
dislocations, and energy of interaction of the disclination directly with
the background impurities, respectively.
In a finite size system with area $A$, we have
\begin{mathletters}
\begin{eqnarray}
E^{\prime}_{D}&=&{\mu \over 72} A\;, \\
E^{\prime}_{BD}&=&s{\mu \pi \over 9} \sum_j (\vec b_j \times \vec r_j)
 \cdot \hat z \ln ({\pi |\vec r_j|^2 \over A})\;,
\label{BD}
\end{eqnarray}
\end{mathletters}
where $\vec b_j$ and $\vec r_j$ are Burger's vectors and the position vectors
of the dislocations,
and $s=\pm 1$ is the ``charge'' of the disclination, specifying whether it is
a 5-fold of 7-fold defect\cite{Nelson}.
The contribution of $\langle E^{\prime}_{D\rho}\rangle$,
where $\langle \dots \rangle$ denotes a disorder average, scales only
as $\sqrt{A\ln A}$, and so is negligible for large system sizes.
We note that $s$ may always be chosen such that
$E^{\prime}_{BD} <0$ for any complexion of dislocations.
The contribution $E^{\prime}_{BD}$ may be estimated by computing the
average over complexions of disclinations.
Assuming for simplicity that the dislocations are completely
uncorrelated,
$\langle \vec b(\vec r) \cdot b(\vec r^\prime) \rangle
\equiv \langle |\vec b| \rangle^2 \delta (\vec r-\vec r^\prime)$,
we find
\begin{equation}
\langle E^{\prime}_{BD}\rangle
 \approx -\sqrt{\langle |E^{\prime}_{BD}|^2 \rangle}
 = -{\langle |\vec b|\rangle \sqrt{\pi} \over {36 a_0}} \mu A\;.
\end{equation}
Thus, the net energy of an isolated disclination for large $A$ is
\begin{equation}
E^{\prime} \approx
[{1 \over 72} - {\langle |\vec b|\rangle \sqrt{\pi} \over {36 a_0}}]
\mu A \;,
\end{equation}
so it is energetically favorable on average to create a disclination if
\begin{equation}
\langle |\vec b|\rangle > {1 \over 2 \sqrt{\pi}}a_0\;.
\end{equation}
Interpreting $\langle |\vec b| \rangle^2$ as the density of dislocations
in lattice units,
we estimate the critical density of dislocations as
$n_b \approx (1/4 \pi) a_0^{-2}$
above which disclinations are energetically favorable.
This value is roughly in agreement with what we observe in our simulations,
in which isolated disclinations appear when $n_b^{-1} \approx$ 17--22 $a_0^{2}$.
We note, finally, that our computation of $\langle E^\prime \rangle$
essentially estimates the mean value of the disclination energy,
if we computed the distribution of $E^\prime$ for all
possible disclination locations.
This mean scales to $\pm \infty$ as $A \propto N_b \to \infty$,
where $N_b$ is the number of dislocations.
One can show that for a simple Gaussian distribution of $\vec b(\vec r)$,
the variance of $\langle E^\prime \rangle$ is also proportional to $N_b$,
so that this model predicts a crossover behavior from a
state with very few disclinations to one with many at the critical
dislocation density.  It is interesting to speculate that a more realistic
distribution for $\vec b(\vec r)$ could convert this crossover to a true
phase transition.  This would be quite consistent with our simulations,
where isolated disclinations are observed for small $d/a_0$, and
there is an apparantly diverging orientational correlation length just
before they disappear.
Details of this calculation will be presented elsewhere\cite{Cha1994c}.

In conclusion, we have observed a crossover between a weak disorder regime
and a strong disorder regime in a study of a model Wigner crystal
under the influence of the modulation-doped donor impurities
as we tune the setback distance.
In the weak disorder regime, we have a hexatic phase where a quasi-long
range orientational order is present, whereas in the strong disorder regime
isolated disclinations destroy the orientational order.
The crossover places $d/a_0 = 1.15 \pm 0.1$.
Associated with this crossover, the behavior of the threshold electric field
changes.
We argue by a continuum elasticity theory
that this crossover can be understood as a proliferation of
disclinations when the density of dislocations is bigger than a certain
critical value.

\acknowledgements
MCC is grateful for helpful discussions with S.~M.~Girvin
and a computational assistance of Clayton Heller.
This work was supported in part by NSF DMR 92-02255.

\begin{figure}
\caption{The orientational correlation functions for various setback distances.
Different symbols represent data for samples with different setback distance.
From top to bottom, $d/a_0=$2.0, 1.7, 1.5, 1.3, 1.2, 1.1, 1.0, 0.9, and 0.8.}
\label{correlation}
\end{figure}

\begin{figure}
\caption{The depinning threshold field observed in various samples
in the unit of $E_0=e/(\epsilon a_0^2)$.
The dotted lines are guides to the eye.}
\label{efield}
\end{figure}

\begin{figure}
\caption{The magnitude of the structure factor, $|S(\vec G)|$,
in the reciprocal lattice vector space for samples with $d/a_0=$
(a)0.8, (b)1.0, (c)1.1, (d)1.2, (e)1.4, and (f)1.7.
Only the points at which $|S(\vec G)| > {1 \over 2}|S(\vec G)|_{\rm max}$
are plotted.
For large setback distances, a six-fold symmetry appears, indicating 
the presence of the quasi-long range orientational order.
}
\label{sfactor}
\end{figure}

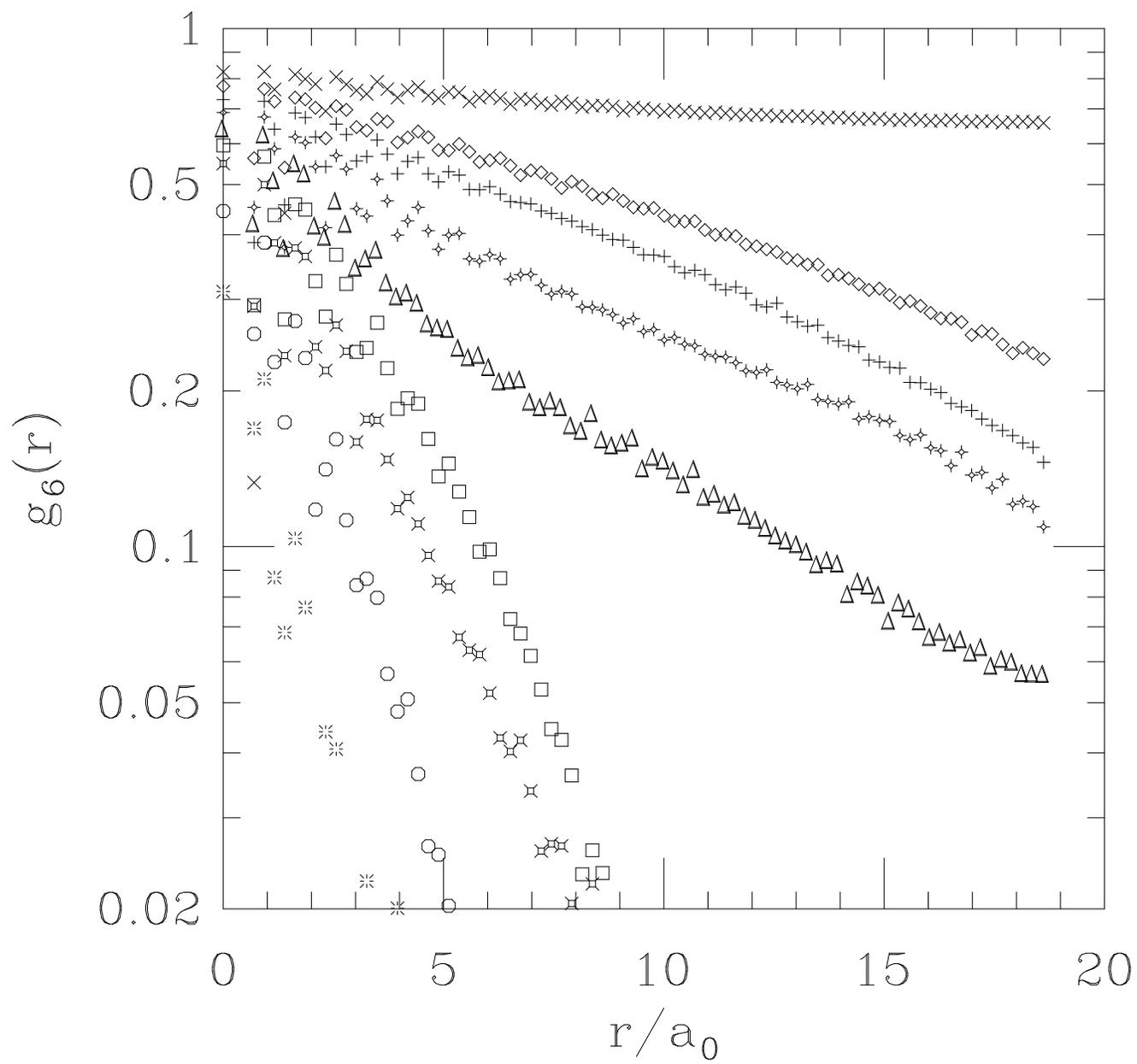
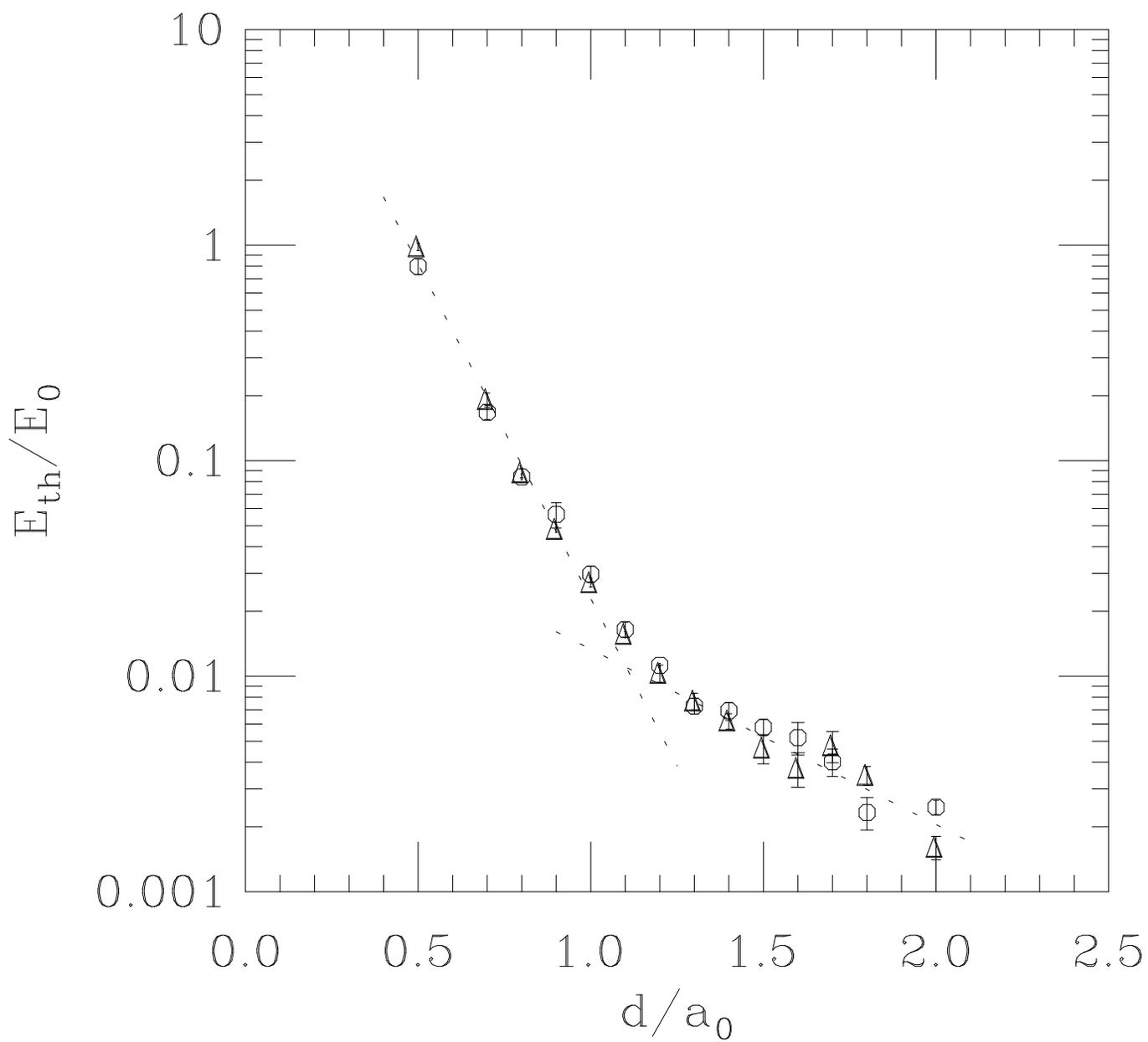
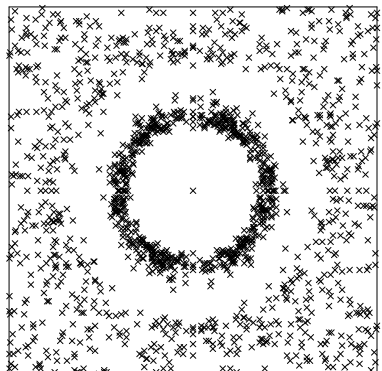
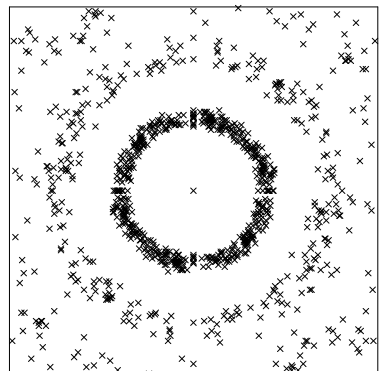
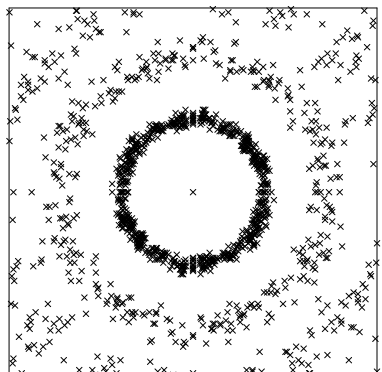
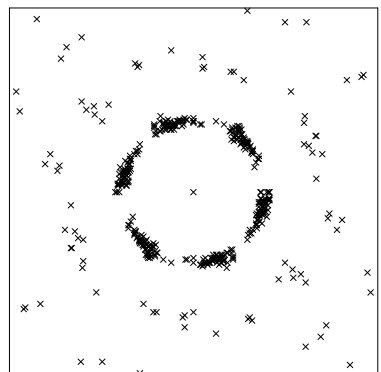
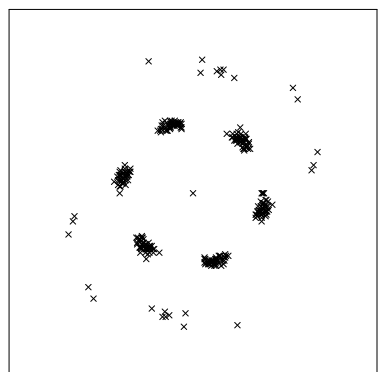
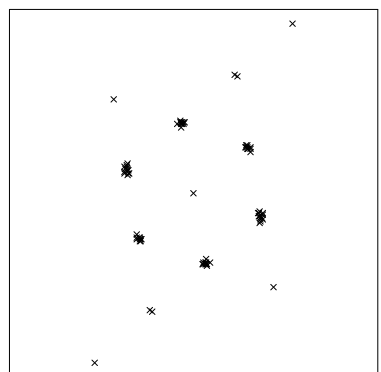
\end{document}